%% file: SemiconductorcQED.tex
\documentclass[%
 aps,
 prl,
 amsmath,amssymb,
reprint,%
showpacs, 
]{revtex4-1}
\usepackage[pdftex]{graphicx}

\begin{document}

\title{Circuit QED using a semiconductor double quantum dot}

\author{H.~Toida}
 \email{toida@thz.c.u-tokyo.ac.jp}
\author{T.~Nakajima}

\author{S.~Komiyama}

\affiliation{Department of Basic Science, The University of Tokyo, Tokyo, Japan}


\begin{abstract}
Vacuum Rabi splitting is observed in a coupled qubit-resonator system consisting of a GaAs double quantum dot and a coplanar waveguide resonator. 
Derived values of the qubit-resonator coupling strength $g$ and the decoherence rate $\Gamma$ indicate strong coupling ($g/\Gamma \sim 2$),  which assures distinct vacuum Rabi oscillation in the system. 
The amplitude of decoherence is reasonably interpreted in terms of the coupling of electrons to piezoelectric acoustic phonons in GaAs. 
\end{abstract}

\maketitle

Hybridizing different quantum systems has attracted considerable attention for quantum information processing and quantum communication, because such systems can utilize the merits of different quantum devices as manipulation capability or long coherence time. 
For this sake coherent coupling between different qubits is essential. 
Superconducting qubits \cite{Sillanpaa2007, Majer2007, Brennen2003} have been strongly coupled to on-chip superconducting resonators forming quantum buses (circuit QED). 
Recent theoretical works \cite{Childress2004, Guo2008, Lin2008, Cottet2010, Jin2011a} suggest that semiconductor quantum dots (QDs) are important and attractive building blocks for circuit QED, because QDs have high scalability, controllability and accessibility to the spin degree of freedom. 
The qubit-resonator coupling strength $g$ and the decoherence rate $\Gamma$ of the system are important two parameters in such systems: 
In a strong coupling regime ($g/\Gamma > 1$), a qubit and a photon are no longer independent physical entities but are coupled into a dressed-atom, which leads to fascinating effects like single atom lasing \cite{Astafiev2007}. 
Recently, dipole coupling between a double quantum dot (DQD) and a superconducting resonator is studied \cite{Frey2012}. 
Real implementation of DQD-based circuit QED, however, is an experimental challenge because of a relatively high decoherence rate. 
Here we report the observation of vacuum Rabi splitting, an indicative of strong coupling between a GaAs/AlGaAs DQD and a superconducting coplanar waveguide (CPW) resonator. 
We also report that the decoherence of the system is dominated by intrinsic piezoelectric effect. 
This work gives a guideline to integrate semiconductor qubits into quantum buses and paves the way towards construction of solid-state hybrid devices for quantum information processing and quantum communication. 

\begin{figure}[htbp]
\includegraphics[width = 0.4\textwidth, clip]{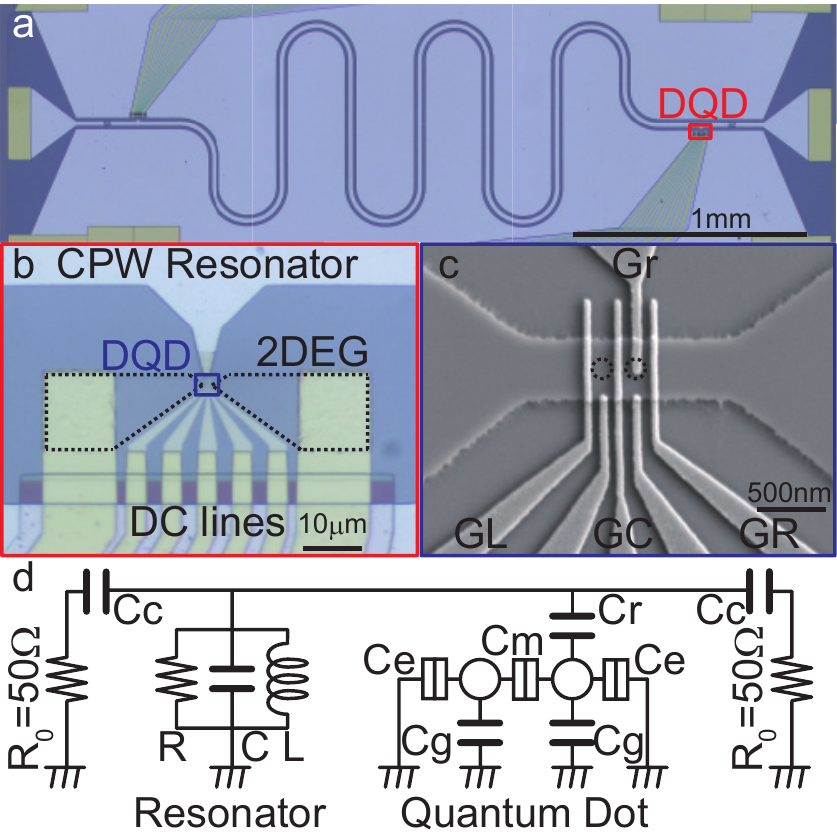}
\caption{(color online). 
(a) Optical microscope image of the coupled DQD-CPW resonator system. The red square marks the location of the DQD, where (b) is the blowup of the region. (c) is a scanning electron microscope image of the DQD defined by biasing metal (NiCr/Au) gates, GL, GC and GR. The right QD is coupled to the resonator with gate electrode Gr via capacitive coupling $C_r \sim$ 50 aF. (d) Equivalent circuit of the device. The two QDs are supposed to be equal except coupling to Gr. The resonator is weakly coupled to the transmission line through $C_c$ = 3 fF. }
\label{fig:sample}
\end{figure}
The device is shown in Figs. \ref{fig:sample} (a) - (c) along with its equivalent circuit in Fig. \ref{fig:sample} (d). 
A DQD (Fig. \ref{fig:sample} (c)) is formed in a GaAs/AlGaAs heterostructure crystal containing a two dimensional electron system (2DES) at a 90nm depth from the surface, where the electron mobility and the sheet electron density are $90 \mathrm{m^2/Vs}$ and $2.3 \times 10^{15}~\mathrm{m^{-2}}$, respectively, at 4.2K. 
A superconducting CPW resonator (Fig. \ref{fig:sample} (a)) is prepared by depositing a 200 nm thick Aluminum layer on top of the crystal surface where the 2DES is removed with 40nm-deep wet etching. 
The resonance frequency and the decay rate are $\omega _0/2\pi$ = 8.3267 GHz and $\kappa /2\pi$ = 8.0 MHz, respectively, at base temperature $\sim$ 30 mK (in the condition when the DQD is not formed). 
The resonator is capacitively coupled to the DQD ($C_r$ see Figs. \ref{fig:sample} (c), (d)). 
Conductance through the DQD is studied via standard lock-in technique. 
The microwave transmitted through the resonator is amplified with a cryogenic amplifier and studied with a vector network analyzer \footnote{
See supplementary material.}. 
All the measurements are made in a dilution refrigerator with a base temperature below 40 mK. 
Thermal excitation of microwave photons in the resonator is completely neglected at 40 mK, with an estimated photon number being less than $5 \times 10^{-5}$. 

\begin{figure*}[htbp]
\includegraphics[width = 0.7\textwidth, clip, viewport=122.0 342.648 470.743 498.0]{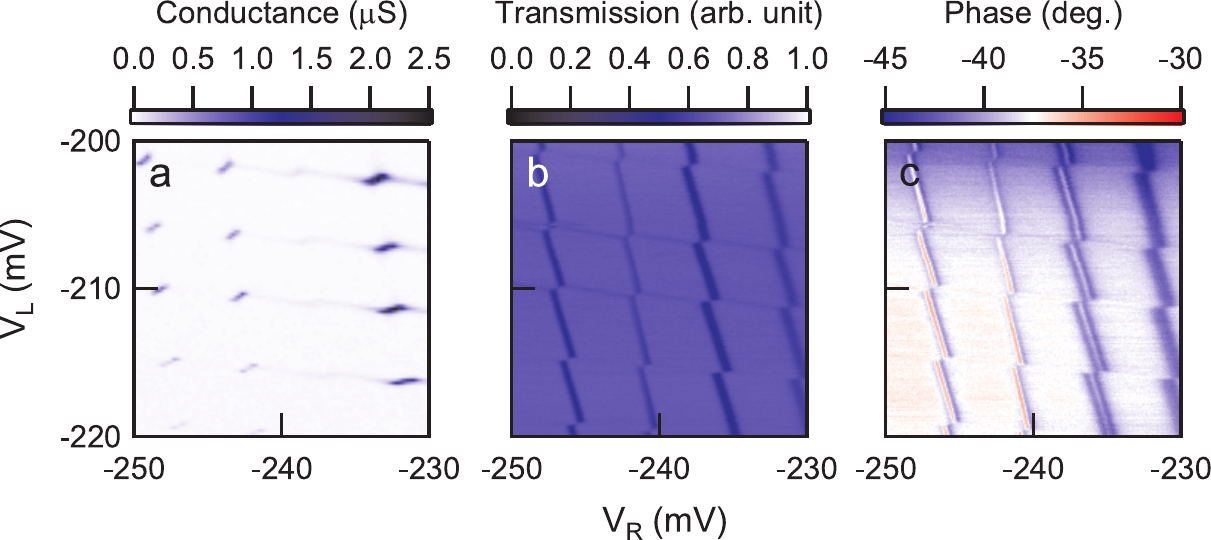}
\caption{(color online).
Charge state of the DQD measured through 
(a) conductance, 
(b) transmission amplitude and 
(c) phase
as a function of $V_L$ and $V_R$. 
$V_C$ is set to -215 mV. 
Transmission amplitude and phase are measured by applying microwave signal at a fixed resonance frequency of 8.3267 GHz. 
}
\label{fig:classical}
\end{figure*}
While the conductance through the DQD is finite only in limited regions of charge triple points  (Fig. 2 (a)) \cite{Wiel2003}, transmission amplitude and phase of the resonator exhibit structures along side edges of the honeycombs (Figs. 2 (b), (c)), where the DQD conductance is vanishing. 
Prior to the experiment, photon assisted tunneling (PAT) has been studied 
\footnote{From the experiment of PAT, we derive conversion factor from gate voltage to frequency. }
. 
In the transmission study throughout the present experiments, the microwave power is strongly suppressed to avoid PAT. 
Dipole coupling between the resonator and the DQD is indicated by the honeycomb structures visible in the two dimensional plots of the resonator transmission amplitude (Fig. 2 (b)) and phase (Fig. 2 (c)). 
The DQD is simply modeled by one capacitance $C_{\mathrm{DQD}}$ and resonance frequency of the system is given by $(2\pi \sqrt{L(C+2C_{c}+C_{\mathrm{DQD}})})^{-1}$. 
$C_{\mathrm{DQD}}$ changes with gate voltages $V_R$ and $V_L$ because connection between the reserver and the QD or between the QDs is turned on or off by Coulomb blockade. 
In our device, because only the right QD is connected to the resonator, large frequency shift occurs along the vertical lines of the honeycombs, where the number of electrons in the right QD changes. 
This asymmetry assures dipole coupling between the DQD and the resonator. 

\begin{figure}[htbp]
\includegraphics[width = 0.48\textwidth, clip, viewport=177.0 342.888 417.051 498.0]{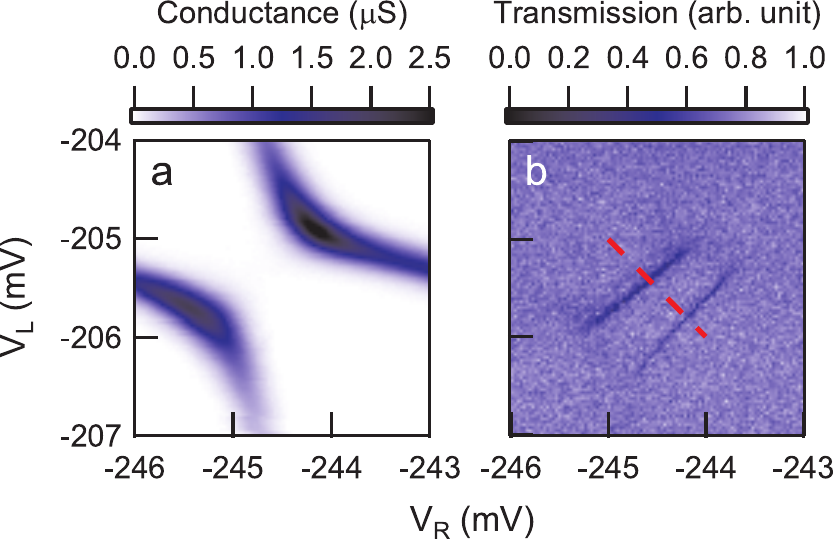}
\caption{(color online). (a) The DQD conductance and (b) the resonator transmission amplitude at 8.3267 GHz in the vicinity of charge triple points, with $V_C$ = -200 mV. 
}
\label{fig:quantum}
\end{figure}
Turning now to the experimental evidence on strong coupling between the DQD and the resonator. 
In the vicinity of the charge triple points, interdot tunneling is allowed and coupling between the DQD and the resonator is expected. 
Conductance through the DQD is discernible only near the charge triple points (Fig. \ref{fig:quantum} (a)). 
Unlike the conductance, remarkably sharp dips in transmission amplitude show up along two parallel lines (Fig. \ref{fig:quantum} (b)). 

\begin{figure}[htbp]
\includegraphics[width = 0.43\textwidth, clip, viewport=200.0 191.561 392.994 649.0]{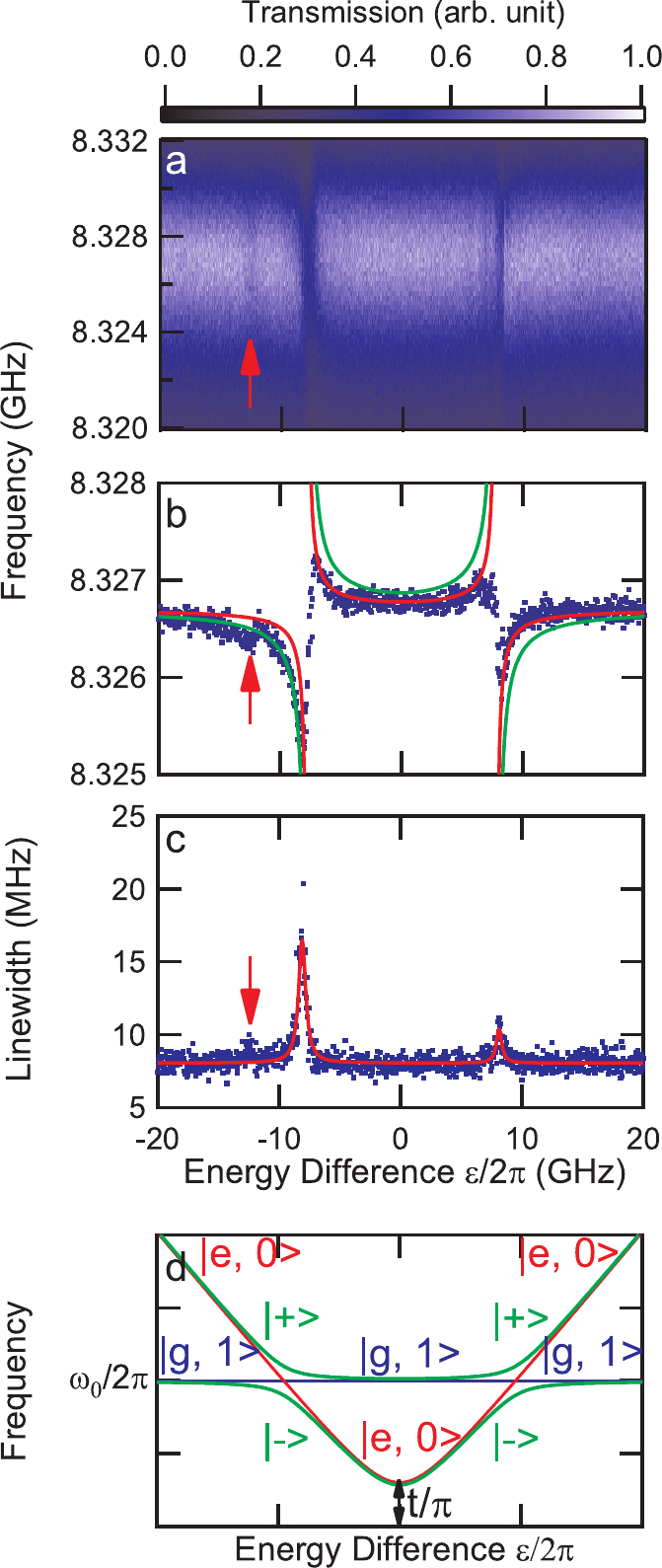}
\caption{(color online). 
(a) Transmission spectra of the resonator as a function of interdot energy difference $\epsilon$. 
$V_L$ and $V_R$ are scanned along the red dashed line in Fig. \ref{fig:quantum} (b).
Gate voltages are converted to energy difference in frequency unit using conversion constant derived from PAT measurements.
(b) The peak position and (c) the linewidth of the transmission spectra. 
The red (Green) line in (b) indicates theoretical values (Eq,(2)) with $g/2\pi = 20~(30)~\mathrm{MHz}$. 
(d) Energy diagram of the DQD (red), the resonator (blue) and the coupled system (green). 
Corresponding quantum states are denoted in the same color. 
}
\label{fig:vrs}
\end{figure}
Next, we measured transmission spectra of the resonator along the red dashed line in Fig. 3 (b).  
Measured spectra (Fig. \ref{fig:vrs} (a)) show two sharp tilted dip lines. 
The peak frequency of the spectra exhibits distinct anticrossings (Fig. \ref{fig:vrs} (b)). 
The resonance linewidth increases significantly in the vicinity of anticrossing points (Fig. \ref{fig:vrs} (c)). 
These remarkable changes occur exactly at the locations where energy difference of the DQD with interdot coupling $\Delta = \sqrt{\epsilon^2+4t^2}$ is in resonance with $\omega _0$ (Fig. \ref{fig:vrs} (d)). 
Here, $t$ is the interdot tunnel rate of the DQD. 
At the anticrossing points, the DQD and the resonator couple via coherent photon exchange. 
As a result, independent description of the DQD and the resonator is replaced by dressed atom state, $|\pm\rangle=(|e, 0\rangle \pm |g, 1\rangle)/\sqrt{2}$ (Fig. \ref{fig:vrs} (d)), where $|g (e), n\rangle$ is the ground (excited) state of the DQD with $n$ photons in the resonator. 
This causes energy splitting corresponding to $|\pm\rangle$ states and anticrossings appear. 

The DQD-resonator coupled  system is described as  
\begin{equation}
	\hat{H} = \frac{\hbar \Delta}{2}\hat{\sigma} _z+\hbar \omega _0 \left( \hat{a}^{\dagger}\hat{a}+ \frac{1}{2} \right)+\hbar g \left( \hat{\sigma} ^{+}\hat{a} + \hat{\sigma} ^{-}\hat{a}^{\dagger}\right)
	\label{eq:JCM}
\end{equation}
according to Jaynes-Cummings model \cite{Childress2004}. 
Here $\hat{\sigma} _{z}$, $\hat{\sigma} _{+}$ and $\hat{\sigma} _{-}$ are the Pauli operators for a DQD, $\hat{a}$ ($\hat{a}^\dagger$) is the annihilation (creation) operator of a photon and $g$ is the coupling strength between the DQD and the resonator. 
The resonance frequency of the coupled system, $\omega _{\pm}$, are given by 
\begin{equation}
	\omega _{\pm} = \frac{\Delta}{2}+\frac{\omega _0}{2} \pm \frac{1}{2}\sqrt{\left(\Delta -\omega _0\right)^2+4g^2}
\end{equation}
when the number of photons in the resonator is much smaller than one. 
The experimental peak frequencies are well reproduced by Eq. (2) with $g/2\pi =20~(30)~\mathrm{MHz}$  for right (left) branch. 
We also estimate decoherence rate of the system $\Gamma /2\pi = 10~(17)~\mathrm{MHz}$ for the right (left) branch from linewidth in Fig. 4 (c). 
Since the system energy is equally distributed between the DQD and the resonator due to vacuum Rabi oscillation at the anticrossing points, $\Gamma$ is represented by the averaged value $\Gamma =(\gamma +\kappa )/2$ of the decohernce rate $\gamma$ due to the DQD and $\kappa$ due to the resonator. 
Combining with independently measured decay rate of the resonator $\kappa /2\pi= 8.0 \mathrm{MHz}$ and $\Gamma$, decoherence rate of the DQD $\gamma /2\pi = 12~(25)~\mathrm{MHz}$ for right (left) branch is also derived. 
With these values, additional parameters give further evaluation of the system \cite{Berman1994}; viz., the number of Rabi oscillation flops $n_{\mathrm{Rabi}}\equiv g/\Gamma =2g/(\kappa +\gamma) = 1.8~(2.0)$, critical photon number $n_0\equiv \gamma ^2/2g^2  = 0.2~(0.3)$ and critical atom number $N_0\equiv2\gamma \kappa/g^2 = 0.5~(0.4)$. 
The physical implication of these parameter values is that two flops of vacuum Rabi oscillation is expected and one photon (atom) is able to bring about saturation in the response of the atom (photon). 
We hence conclude that system is in a strong coupling regime with distinct vacuum Rabi oscillation
\footnote{Additional small dip indicated by the red arrows is ascribed to the excited state of the DQD (Figs. \ref{fig:vrs} (a) - (c)).}.

Decoherence rate of a DQD $\gamma$ is expressed as sum of the energy relaxation rate $\gamma _{1}$ and the dephasing rate $\gamma _{\phi}$: $\gamma = \gamma _{1}/2 + \gamma _{\phi}$ \cite{Wangsness1953}. 
The energy relaxation is dominated by emission of piezoelectric acoustic phonons for GaAs/AlGaAs DQDs \cite{Fujisawa1998, Brandes1999}, which is intrinsic property of GaAs. 
Pulsed gate measurement of a DQD \cite{Hayashi2003} give an estimation of $\gamma_1/2\pi \sim 8 - 40~\mathrm{MHz}$ due to piezoelectric acoustic phonon interaction. 
These values are comparable to our $\gamma /2\pi = 12~(25)~\mathrm{MHz}$.
For our device, dephasing rate $\gamma _{\phi}/2\pi < \gamma/2\pi =12~(25)~\mathrm{MHz}$ is small enough to distinguish vacuum Rabi splitting clearly. 
In contrast, anticrossings are not distinguished in earlier experiment \cite{Frey2012}, because dephasing rate is much larger than coupling strength, $\gamma _{\phi} \gg g$. 
Dephasing rate of gated GaAs/AlGaAs devices is dominated by device specific background charge fluctuation \cite{Hayashi2003, Petersson2010b} and can be reduced by bias cooling method \cite{Buizert2008}. 

Energy loss of the resonator $\kappa /2\pi = 8.0~\mathrm{MHz}$ occurs through either the photon escape to external transmission line (coupling loss) or the photon loss inside the resonator (internal loss) \cite{Frunzio2005}. 
For our device, coupling capacitance $C_{c}=3~\mathrm{fF}$ is so small that the coupling loss $\kappa_{\mathrm{ext}} /2\pi = 0.4~\mathrm{MHz}$ is ignored compared to the internal loss. 
Internal loss can arise from radiative loss, metallic loss and dielectric loss. 
For the CPW resonator used in this experiment, radiative loss and metallic loss is suppressed by designing typical size of the resonator to be much smaller than wavelength and using superconductor. 
Indeed, the decay rate of CPWs is reported to be dependent on the substrate material, where the order of rate is Si $<$ sapphire $<$ Si$\mathrm{N}_{\mathrm{x}}$ $<$ Si$\mathrm{O}_2$ $<$ AlN $<$ MgO \cite{O'Connell2008}. 
The present authors suggest that the material specific feature along with the absolute amplitude of loss is ascribed to the piezoelectric effect; that is, CPW resonators produce oscillating electrical fields in a substrate between ground plane and center conductor, which generates dissipative lattice vibration via piezoelectric interaction. 
From piezoelectric constant of GaAs $d_{14} = 2.7~\mathrm{pC/N}$ \cite{Sadao1985}, the decay rate of the resonator is estimated to be 1.6 - 16 MHz which is comparable to our experimental value $\kappa /2\pi = 8.0~\mathrm{MHz}$. 

For GaAs based circuit QED devices, decoherence rate of the DQD and the CPW resonator is dominated by piezoelectric effect, which is intrinsic property of GaAs.  
For our device, decoherence rate is small enough to observe strong coupling peculiar features. 
Further improvement of coupling strength to the decoherence rate ratio, $g/\Gamma$, is possible by using less piezo-active materials such as Si, SiGe, carbon nanotube and graphene. 

In conclusion, we have realized strongly coupled system using a DQD and a CPW resonator. 
System decoherence rate, which is directly measured through vacuum Rabi splitting, suggest that the decoherence mechanism of our system is dominated by intrinsic piezoelectric effect, which will be improved by using less piezo-active materials. 
\begin{acknowledgments}
This work is supported by KAKENHI, Grant-in-Aid for JSPS Fellows, No. 22-861.
We thank for fruitful discussion with T. Fujisawa and O. Asterfiev.
\end{acknowledgments}

\input{SemiconductorcQED.bbl}

\end{document}

%% file: SemiconductorcQED.bbl
%

%% file: SemiconductorcQED.bbl
\begin{thebibliography}{24}%
\makeatletter
\providecommand \@ifxundefined [1]{%
 \@ifx{#1\undefined}
}%
\providecommand \@ifnum [1]{%
 \ifnum #1\expandafter \@firstoftwo
 \else \expandafter \@secondoftwo
 \fi
}%
\providecommand \@ifx [1]{%
 \ifx #1\expandafter \@firstoftwo
 \else \expandafter \@secondoftwo
 \fi
}%
\providecommand \natexlab [1]{#1}%
\providecommand \enquote  [1]{``#1''}%
\providecommand \bibnamefont  [1]{#1}%
\providecommand \bibfnamefont [1]{#1}%
\providecommand \citenamefont [1]{#1}%
\providecommand \href@noop [0]{\@secondoftwo}%
\providecommand \href [0]{\begingroup \@sanitize@url \@href}%
\providecommand \@href[1]{\@@startlink{#1}\@@href}%
\providecommand \@@href[1]{\endgroup#1\@@endlink}%
\providecommand \@sanitize@url [0]{\catcode `\\12\catcode `\$12\catcode
  `\&12\catcode `\#12\catcode `\^12\catcode `\_12\catcode `\%12\relax}%
\providecommand \@@startlink[1]{}%
\providecommand \@@endlink[0]{}%
\providecommand \url  [0]{\begingroup\@sanitize@url \@url }%
\providecommand \@url [1]{\endgroup\@href {#1}{\urlprefix }}%
\providecommand \urlprefix  [0]{URL }%
\providecommand \Eprint [0]{\href }%
\providecommand \doibase [0]{http://dx.doi.org/}%
\providecommand \selectlanguage [0]{\@gobble}%
\providecommand \bibinfo  [0]{\@secondoftwo}%
\providecommand \bibfield  [0]{\@secondoftwo}%
\providecommand \translation [1]{[#1]}%
\providecommand \BibitemOpen [0]{}%
\providecommand \bibitemStop [0]{}%
\providecommand \bibitemNoStop [0]{.\EOS\space}%
\providecommand \EOS [0]{\spacefactor3000\relax}%
\providecommand \BibitemShut  [1]{\csname bibitem#1\endcsname}%
\let\auto@bib@innerbib\@empty
\bibitem [{\citenamefont {Sillanpaa}\ \emph {et~al.}(2007)\citenamefont
  {Sillanpaa}, \citenamefont {Park},\ and\ \citenamefont
  {Simmonds}}]{Sillanpaa2007}%
  \BibitemOpen
  \bibfield  {author} {\bibinfo {author} {\bibfnamefont {M.~A.}\ \bibnamefont
  {Sillanpaa}}, \bibinfo {author} {\bibfnamefont {J.~I.}\ \bibnamefont {Park}},
  \ and\ \bibinfo {author} {\bibfnamefont {R.~W.}\ \bibnamefont {Simmonds}},\
  }\href {http://dx.doi.org/10.1038/nature06124} {\bibfield  {journal}
  {\bibinfo  {journal} {Nature}\ }\textbf {\bibinfo {volume} {449}},\ \bibinfo
  {pages} {438} (\bibinfo {year} {2007})}\BibitemShut {NoStop}%
\bibitem [{\citenamefont {Majer}\ \emph {et~al.}(2007)\citenamefont {Majer},
  \citenamefont {Chow}, \citenamefont {Gambetta}, \citenamefont {Koch},
  \citenamefont {Johnson}, \citenamefont {Schreier}, \citenamefont {Frunzio},
  \citenamefont {Schuster}, \citenamefont {Houck}, \citenamefont {Wallraff},
  \citenamefont {Blais}, \citenamefont {Devoret}, \citenamefont {Girvin},\ and\
  \citenamefont {Schoelkopf}}]{Majer2007}%
  \BibitemOpen
  \bibfield  {author} {\bibinfo {author} {\bibfnamefont {J.}~\bibnamefont
  {Majer}}, \bibinfo {author} {\bibfnamefont {J.~M.}\ \bibnamefont {Chow}},
  \bibinfo {author} {\bibfnamefont {J.~M.}\ \bibnamefont {Gambetta}}, \bibinfo
  {author} {\bibfnamefont {J.}~\bibnamefont {Koch}}, \bibinfo {author}
  {\bibfnamefont {B.~R.}\ \bibnamefont {Johnson}}, \bibinfo {author}
  {\bibfnamefont {J.~A.}\ \bibnamefont {Schreier}}, \bibinfo {author}
  {\bibfnamefont {L.}~\bibnamefont {Frunzio}}, \bibinfo {author} {\bibfnamefont
  {D.~I.}\ \bibnamefont {Schuster}}, \bibinfo {author} {\bibfnamefont {A.~A.}\
  \bibnamefont {Houck}}, \bibinfo {author} {\bibfnamefont {A.}~\bibnamefont
  {Wallraff}}, \bibinfo {author} {\bibfnamefont {A.}~\bibnamefont {Blais}},
  \bibinfo {author} {\bibfnamefont {M.~H.}\ \bibnamefont {Devoret}}, \bibinfo
  {author} {\bibfnamefont {S.~M.}\ \bibnamefont {Girvin}}, \ and\ \bibinfo
  {author} {\bibfnamefont {R.~J.}\ \bibnamefont {Schoelkopf}},\ }\href
  {http://dx.doi.org/10.1038/nature06184} {\bibfield  {journal} {\bibinfo
  {journal} {Nature}\ }\textbf {\bibinfo {volume} {449}},\ \bibinfo {pages}
  {443} (\bibinfo {year} {2007})}\BibitemShut {NoStop}%
\bibitem [{\citenamefont {Brennen}\ \emph {et~al.}(2003)\citenamefont
  {Brennen}, \citenamefont {Song},\ and\ \citenamefont
  {Williams}}]{Brennen2003}%
  \BibitemOpen
  \bibfield  {author} {\bibinfo {author} {\bibfnamefont {G.~K.}\ \bibnamefont
  {Brennen}}, \bibinfo {author} {\bibfnamefont {D.}~\bibnamefont {Song}}, \
  and\ \bibinfo {author} {\bibfnamefont {C.~J.}\ \bibnamefont {Williams}},\
  }\href {http://link.aps.org/doi/10.1103/PhysRevA.67.050302} {\bibfield
  {journal} {\bibinfo  {journal} {Physical Review A}\ }\textbf {\bibinfo
  {volume} {67}},\ \bibinfo {pages} {050302} (\bibinfo {year}
  {2003})}\BibitemShut {NoStop}%
\bibitem [{\citenamefont {Childress}\ \emph {et~al.}(2004)\citenamefont
  {Childress}, \citenamefont {S{\o}rensen},\ and\ \citenamefont
  {Lukin}}]{Childress2004}%
  \BibitemOpen
  \bibfield  {author} {\bibinfo {author} {\bibfnamefont {L.}~\bibnamefont
  {Childress}}, \bibinfo {author} {\bibfnamefont {A.~S.}\ \bibnamefont
  {S{\o}rensen}}, \ and\ \bibinfo {author} {\bibfnamefont {M.~D.}\ \bibnamefont
  {Lukin}},\ }\href {http://link.aps.org/doi/10.1103/PhysRevA.69.042302}
  {\bibfield  {journal} {\bibinfo  {journal} {Physical Review A}\ }\textbf
  {\bibinfo {volume} {69}},\ \bibinfo {pages} {042302} (\bibinfo {year}
  {2004})}\BibitemShut {NoStop}%
\bibitem [{\citenamefont {Guo}\ \emph {et~al.}(2008)\citenamefont {Guo},
  \citenamefont {Zhang}, \citenamefont {Hu}, \citenamefont {Tu},\ and\
  \citenamefont {Guo}}]{Guo2008}%
  \BibitemOpen
  \bibfield  {author} {\bibinfo {author} {\bibfnamefont {G.-P.}\ \bibnamefont
  {Guo}}, \bibinfo {author} {\bibfnamefont {H.}~\bibnamefont {Zhang}}, \bibinfo
  {author} {\bibfnamefont {Y.}~\bibnamefont {Hu}}, \bibinfo {author}
  {\bibfnamefont {T.}~\bibnamefont {Tu}}, \ and\ \bibinfo {author}
  {\bibfnamefont {G.-C.}\ \bibnamefont {Guo}},\ }\href
  {http://link.aps.org/doi/10.1103/PhysRevA.78.020302} {\bibfield  {journal}
  {\bibinfo  {journal} {Physical Review A}\ }\textbf {\bibinfo {volume} {78}},\
  \bibinfo {pages} {020302} (\bibinfo {year} {2008})}\BibitemShut {NoStop}%
\bibitem [{\citenamefont {Lin}\ \emph {et~al.}(2008)\citenamefont {Lin},
  \citenamefont {Guo}, \citenamefont {Tu}, \citenamefont {Zhu},\ and\
  \citenamefont {Guo}}]{Lin2008}%
  \BibitemOpen
  \bibfield  {author} {\bibinfo {author} {\bibfnamefont {Z.-R.}\ \bibnamefont
  {Lin}}, \bibinfo {author} {\bibfnamefont {G.-P.}\ \bibnamefont {Guo}},
  \bibinfo {author} {\bibfnamefont {T.}~\bibnamefont {Tu}}, \bibinfo {author}
  {\bibfnamefont {F.-Y.}\ \bibnamefont {Zhu}}, \ and\ \bibinfo {author}
  {\bibfnamefont {G.-C.}\ \bibnamefont {Guo}},\ }\href
  {http://link.aps.org/doi/10.1103/PhysRevLett.101.230501} {\bibfield
  {journal} {\bibinfo  {journal} {Physical Review Letters}\ }\textbf {\bibinfo
  {volume} {101}},\ \bibinfo {pages} {230501} (\bibinfo {year}
  {2008})}\BibitemShut {NoStop}%
\bibitem [{\citenamefont {Cottet}\ and\ \citenamefont
  {Kontos}(2010)}]{Cottet2010}%
  \BibitemOpen
  \bibfield  {author} {\bibinfo {author} {\bibfnamefont {A.}~\bibnamefont
  {Cottet}}\ and\ \bibinfo {author} {\bibfnamefont {T.}~\bibnamefont
  {Kontos}},\ }\href {http://link.aps.org/doi/10.1103/PhysRevLett.105.160502}
  {\bibfield  {journal} {\bibinfo  {journal} {Physical Review Letters}\
  }\textbf {\bibinfo {volume} {105}},\ \bibinfo {pages} {160502} (\bibinfo
  {year} {2010})}\BibitemShut {NoStop}%
\bibitem [{\citenamefont {Jin}\ \emph {et~al.}(2011)\citenamefont {Jin},
  \citenamefont {Marthaler}, \citenamefont {Cole}, \citenamefont {Shnirman},\
  and\ \citenamefont {Sch\"on}}]{Jin2011a}%
  \BibitemOpen
  \bibfield  {author} {\bibinfo {author} {\bibfnamefont {P.-Q.}\ \bibnamefont
  {Jin}}, \bibinfo {author} {\bibfnamefont {M.}~\bibnamefont {Marthaler}},
  \bibinfo {author} {\bibfnamefont {J.~H.}\ \bibnamefont {Cole}}, \bibinfo
  {author} {\bibfnamefont {A.}~\bibnamefont {Shnirman}}, \ and\ \bibinfo
  {author} {\bibfnamefont {G.}~\bibnamefont {Sch\"on}},\ }\href
  {http://link.aps.org/doi/10.1103/PhysRevB.84.035322} {\bibfield  {journal}
  {\bibinfo  {journal} {Physical Review B}\ }\textbf {\bibinfo {volume} {84}},\
  \bibinfo {pages} {035322} (\bibinfo {year} {2011})}\BibitemShut {NoStop}%
\bibitem [{\citenamefont {Astafiev}\ \emph {et~al.}(2007)\citenamefont
  {Astafiev}, \citenamefont {Inomata}, \citenamefont {Niskanen}, \citenamefont
  {Yamamoto}, \citenamefont {Pashkin}, \citenamefont {Nakamura},\ and\
  \citenamefont {Tsai}}]{Astafiev2007}%
  \BibitemOpen
  \bibfield  {author} {\bibinfo {author} {\bibfnamefont {O.}~\bibnamefont
  {Astafiev}}, \bibinfo {author} {\bibfnamefont {K.}~\bibnamefont {Inomata}},
  \bibinfo {author} {\bibfnamefont {A.~O.}\ \bibnamefont {Niskanen}}, \bibinfo
  {author} {\bibfnamefont {T.}~\bibnamefont {Yamamoto}}, \bibinfo {author}
  {\bibfnamefont {Y.~A.}\ \bibnamefont {Pashkin}}, \bibinfo {author}
  {\bibfnamefont {Y.}~\bibnamefont {Nakamura}}, \ and\ \bibinfo {author}
  {\bibfnamefont {J.~S.}\ \bibnamefont {Tsai}},\ }\href
  {http://dx.doi.org/10.1038/nature06141} {\bibfield  {journal} {\bibinfo
  {journal} {Nature}\ }\textbf {\bibinfo {volume} {449}},\ \bibinfo {pages}
  {588} (\bibinfo {year} {2007})}\BibitemShut {NoStop}%
\bibitem [{\citenamefont {Frey}\ \emph {et~al.}(2012)\citenamefont {Frey},
  \citenamefont {Leek}, \citenamefont {Beck}, \citenamefont {Blais},
  \citenamefont {Ihn}, \citenamefont {Ensslin},\ and\ \citenamefont
  {Wallraff}}]{Frey2012}%
  \BibitemOpen
  \bibfield  {author} {\bibinfo {author} {\bibfnamefont {T.}~\bibnamefont
  {Frey}}, \bibinfo {author} {\bibfnamefont {P.~J.}\ \bibnamefont {Leek}},
  \bibinfo {author} {\bibfnamefont {M.}~\bibnamefont {Beck}}, \bibinfo {author}
  {\bibfnamefont {A.}~\bibnamefont {Blais}}, \bibinfo {author} {\bibfnamefont
  {T.}~\bibnamefont {Ihn}}, \bibinfo {author} {\bibfnamefont {K.}~\bibnamefont
  {Ensslin}}, \ and\ \bibinfo {author} {\bibfnamefont {A.}~\bibnamefont
  {Wallraff}},\ }\href {http://link.aps.org/doi/10.1103/PhysRevLett.108.046807}
  {\bibfield  {journal} {\bibinfo  {journal} {Physical Review Letters}\
  }\textbf {\bibinfo {volume} {108}},\ \bibinfo {pages} {046807} (\bibinfo
  {year} {2012})}\BibitemShut {NoStop}%
\bibitem [{Note1()}]{Note1}%
  \BibitemOpen
  \bibinfo {note} {See supplementary material.}\BibitemShut {Stop}%
\bibitem [{\citenamefont {van~der Wiel}\ \emph {et~al.}(2003)\citenamefont
  {van~der Wiel}, \citenamefont {De~Franceschi}, \citenamefont {Elzerman},
  \citenamefont {Fujisawa}, \citenamefont {Tarucha},\ and\ \citenamefont
  {Kouwenhoven}}]{Wiel2003}%
  \BibitemOpen
  \bibfield  {author} {\bibinfo {author} {\bibfnamefont {W.~G.}\ \bibnamefont
  {van~der Wiel}}, \bibinfo {author} {\bibfnamefont {S.}~\bibnamefont
  {De~Franceschi}}, \bibinfo {author} {\bibfnamefont {J.~M.}\ \bibnamefont
  {Elzerman}}, \bibinfo {author} {\bibfnamefont {T.}~\bibnamefont {Fujisawa}},
  \bibinfo {author} {\bibfnamefont {S.}~\bibnamefont {Tarucha}}, \ and\
  \bibinfo {author} {\bibfnamefont {L.~P.}\ \bibnamefont {Kouwenhoven}},\
  }\href {http://link.aps.org/doi/10.1103/RevModPhys.75.1} {\bibfield
  {journal} {\bibinfo  {journal} {Reviews of Modern Physics}\ }\textbf
  {\bibinfo {volume} {75}},\ \bibinfo {pages} {1} (\bibinfo {year}
  {2003})}\BibitemShut {NoStop}%
\bibitem [{Note2()}]{Note2}%
  \BibitemOpen
  \bibinfo {note} {From the experiment of PAT, we derive conversion factor from
  gate voltage to frequency.}\BibitemShut {Stop}%
\bibitem [{\citenamefont {Berman}(1994)}]{Berman1994}%
  \BibitemOpen
  \bibinfo {editor} {\bibfnamefont {P.~R.}\ \bibnamefont {Berman}},\ ed.,\
  \href@noop {} {\emph {\bibinfo {title} {Cavity Quantum Electrodynamics}}}\
  (\bibinfo  {publisher} {Academic Press},\ \bibinfo {address} {San Diego},\
  \bibinfo {year} {1994})\BibitemShut {NoStop}%
\bibitem [{Note3()}]{Note3}%
  \BibitemOpen
  \bibinfo {note} {Additional small dip indicated by the red arrows is ascribed
  to the excited state of the DQD (Figs. \ref {fig:vrs} (a) -
  (c)).}\BibitemShut {Stop}%
\bibitem [{\citenamefont {Wangsness}\ and\ \citenamefont
  {Bloch}(1953)}]{Wangsness1953}%
  \BibitemOpen
  \bibfield  {author} {\bibinfo {author} {\bibfnamefont {R.~K.}\ \bibnamefont
  {Wangsness}}\ and\ \bibinfo {author} {\bibfnamefont {F.}~\bibnamefont
  {Bloch}},\ }\href {http://link.aps.org/doi/10.1103/PhysRev.89.728} {\bibfield
   {journal} {\bibinfo  {journal} {Physical Review}\ }\textbf {\bibinfo
  {volume} {89}},\ \bibinfo {pages} {728} (\bibinfo {year} {1953})}\BibitemShut
  {NoStop}%
\bibitem [{\citenamefont {Fujisawa}\ \emph {et~al.}(1998)\citenamefont
  {Fujisawa}, \citenamefont {Oosterkamp}, \citenamefont {van~der Wiel},
  \citenamefont {Broer}, \citenamefont {Aguado}, \citenamefont {Tarucha},\ and\
  \citenamefont {Kouwenhoven}}]{Fujisawa1998}%
  \BibitemOpen
  \bibfield  {author} {\bibinfo {author} {\bibfnamefont {T.}~\bibnamefont
  {Fujisawa}}, \bibinfo {author} {\bibfnamefont {T.~H.}\ \bibnamefont
  {Oosterkamp}}, \bibinfo {author} {\bibfnamefont {W.~G.}\ \bibnamefont
  {van~der Wiel}}, \bibinfo {author} {\bibfnamefont {B.~W.}\ \bibnamefont
  {Broer}}, \bibinfo {author} {\bibfnamefont {R.}~\bibnamefont {Aguado}},
  \bibinfo {author} {\bibfnamefont {S.}~\bibnamefont {Tarucha}}, \ and\
  \bibinfo {author} {\bibfnamefont {L.~P.}\ \bibnamefont {Kouwenhoven}},\
  }\href {http://www.sciencemag.org/content/282/5390/932.abstract} {\bibfield
  {journal} {\bibinfo  {journal} {Science}\ }\textbf {\bibinfo {volume}
  {282}},\ \bibinfo {pages} {932} (\bibinfo {year} {1998})}\BibitemShut
  {NoStop}%
\bibitem [{\citenamefont {Brandes}\ and\ \citenamefont
  {Kramer}(1999)}]{Brandes1999}%
  \BibitemOpen
  \bibfield  {author} {\bibinfo {author} {\bibfnamefont {T.}~\bibnamefont
  {Brandes}}\ and\ \bibinfo {author} {\bibfnamefont {B.}~\bibnamefont
  {Kramer}},\ }\href {http://link.aps.org/doi/10.1103/PhysRevLett.83.3021}
  {\bibfield  {journal} {\bibinfo  {journal} {Physical Review Letters}\
  }\textbf {\bibinfo {volume} {83}},\ \bibinfo {pages} {3021} (\bibinfo {year}
  {1999})}\BibitemShut {NoStop}%
\bibitem [{\citenamefont {Hayashi}\ \emph {et~al.}(2003)\citenamefont
  {Hayashi}, \citenamefont {Fujisawa}, \citenamefont {Cheong}, \citenamefont
  {Jeong},\ and\ \citenamefont {Hirayama}}]{Hayashi2003}%
  \BibitemOpen
  \bibfield  {author} {\bibinfo {author} {\bibfnamefont {T.}~\bibnamefont
  {Hayashi}}, \bibinfo {author} {\bibfnamefont {T.}~\bibnamefont {Fujisawa}},
  \bibinfo {author} {\bibfnamefont {H.~D.}\ \bibnamefont {Cheong}}, \bibinfo
  {author} {\bibfnamefont {Y.~H.}\ \bibnamefont {Jeong}}, \ and\ \bibinfo
  {author} {\bibfnamefont {Y.}~\bibnamefont {Hirayama}},\ }\href
  {http://link.aps.org/doi/10.1103/PhysRevLett.91.226804} {\bibfield  {journal}
  {\bibinfo  {journal} {Physical Review Letters}\ }\textbf {\bibinfo {volume}
  {91}},\ \bibinfo {pages} {226804} (\bibinfo {year} {2003})}\BibitemShut
  {NoStop}%
\bibitem [{\citenamefont {Petersson}\ \emph {et~al.}(2010)\citenamefont
  {Petersson}, \citenamefont {Petta}, \citenamefont {Lu},\ and\ \citenamefont
  {Gossard}}]{Petersson2010b}%
  \BibitemOpen
  \bibfield  {author} {\bibinfo {author} {\bibfnamefont {K.~D.}\ \bibnamefont
  {Petersson}}, \bibinfo {author} {\bibfnamefont {J.~R.}\ \bibnamefont
  {Petta}}, \bibinfo {author} {\bibfnamefont {H.}~\bibnamefont {Lu}}, \ and\
  \bibinfo {author} {\bibfnamefont {A.~C.}\ \bibnamefont {Gossard}},\ }\href
  {http://link.aps.org/doi/10.1103/PhysRevLett.105.246804} {\bibfield
  {journal} {\bibinfo  {journal} {Physical Review Letters}\ }\textbf {\bibinfo
  {volume} {105}},\ \bibinfo {pages} {246804} (\bibinfo {year}
  {2010})}\BibitemShut {NoStop}%
\bibitem [{\citenamefont {Buizert}\ \emph {et~al.}(2008)\citenamefont
  {Buizert}, \citenamefont {Koppens}, \citenamefont {Pioro-Ladri\'ere},
  \citenamefont {Tranitz}, \citenamefont {Vink}, \citenamefont {Tarucha},
  \citenamefont {Wegscheider},\ and\ \citenamefont
  {Vandersypen}}]{Buizert2008}%
  \BibitemOpen
  \bibfield  {author} {\bibinfo {author} {\bibfnamefont {C.}~\bibnamefont
  {Buizert}}, \bibinfo {author} {\bibfnamefont {F.~H.~L.}\ \bibnamefont
  {Koppens}}, \bibinfo {author} {\bibfnamefont {M.}~\bibnamefont
  {Pioro-Ladri\'ere}, \bibfnamefont {M.re}}, \bibinfo {author} {\bibfnamefont
  {H.-P.}\ \bibnamefont {Tranitz}}, \bibinfo {author} {\bibfnamefont {I.~T.}\
  \bibnamefont {Vink}}, \bibinfo {author} {\bibfnamefont {S.}~\bibnamefont
  {Tarucha}}, \bibinfo {author} {\bibfnamefont {W.}~\bibnamefont
  {Wegscheider}}, \ and\ \bibinfo {author} {\bibfnamefont {L.~M.~K.}\
  \bibnamefont {Vandersypen}},\ }\href
  {http://link.aps.org/doi/10.1103/PhysRevLett.101.226603} {\bibfield
  {journal} {\bibinfo  {journal} {Physical Review Letters}\ }\textbf {\bibinfo
  {volume} {101}},\ \bibinfo {pages} {226603} (\bibinfo {year}
  {2008})}\BibitemShut {NoStop}%
\bibitem [{\citenamefont {Frunzio}\ \emph {et~al.}(2005)\citenamefont
  {Frunzio}, \citenamefont {Wallraff}, \citenamefont {Schuster}, \citenamefont
  {Majer},\ and\ \citenamefont {Schoelkopf}}]{Frunzio2005}%
  \BibitemOpen
  \bibfield  {author} {\bibinfo {author} {\bibfnamefont {L.}~\bibnamefont
  {Frunzio}}, \bibinfo {author} {\bibfnamefont {A.}~\bibnamefont {Wallraff}},
  \bibinfo {author} {\bibfnamefont {D.}~\bibnamefont {Schuster}}, \bibinfo
  {author} {\bibfnamefont {J.}~\bibnamefont {Majer}}, \ and\ \bibinfo {author}
  {\bibfnamefont {R.}~\bibnamefont {Schoelkopf}},\ }\href
  {http://ieeexplore.ieee.org/stamp/stamp.jsp?tp=&arnumber=1439774&isnumber=31006}
  {\bibfield  {journal} {\bibinfo  {journal} {IEEE Transactions on Applied
  Superconductivity}\ }\textbf {\bibinfo {volume} {15}},\ \bibinfo {pages}
  {860} (\bibinfo {year} {2005})}\BibitemShut {NoStop}%
\bibitem [{\citenamefont {O'Connell}\ \emph {et~al.}(2008)\citenamefont
  {O'Connell}, \citenamefont {Ansmann}, \citenamefont {Bialczak}, \citenamefont
  {Hofheinz}, \citenamefont {Katz}, \citenamefont {Lucero}, \citenamefont
  {McKenney}, \citenamefont {Neeley}, \citenamefont {Wang}, \citenamefont
  {Weig}, \citenamefont {Cleland},\ and\ \citenamefont
  {Martinis}}]{O'Connell2008}%
  \BibitemOpen
  \bibfield  {author} {\bibinfo {author} {\bibfnamefont {A.~D.}\ \bibnamefont
  {O'Connell}}, \bibinfo {author} {\bibfnamefont {M.}~\bibnamefont {Ansmann}},
  \bibinfo {author} {\bibfnamefont {R.~C.}\ \bibnamefont {Bialczak}}, \bibinfo
  {author} {\bibfnamefont {M.}~\bibnamefont {Hofheinz}}, \bibinfo {author}
  {\bibfnamefont {N.}~\bibnamefont {Katz}}, \bibinfo {author} {\bibfnamefont
  {E.}~\bibnamefont {Lucero}}, \bibinfo {author} {\bibfnamefont
  {C.}~\bibnamefont {McKenney}}, \bibinfo {author} {\bibfnamefont
  {M.}~\bibnamefont {Neeley}}, \bibinfo {author} {\bibfnamefont
  {H.}~\bibnamefont {Wang}}, \bibinfo {author} {\bibfnamefont {E.~M.}\
  \bibnamefont {Weig}}, \bibinfo {author} {\bibfnamefont {A.~N.}\ \bibnamefont
  {Cleland}}, \ and\ \bibinfo {author} {\bibfnamefont {J.~M.}\ \bibnamefont
  {Martinis}},\ }\href {http://link.aip.org/link/?APL/92/112903/1} {\bibfield
  {journal} {\bibinfo  {journal} {Applied Physics Letters}\ }\textbf {\bibinfo
  {volume} {92}},\ \bibinfo {pages} {112903} (\bibinfo {year}
  {2008})}\BibitemShut {NoStop}%
\bibitem [{\citenamefont {Sadao}(1985)}]{Sadao1985}%
  \BibitemOpen
  \bibfield  {author} {\bibinfo {author} {\bibfnamefont {A.}~\bibnamefont
  {Sadao}},\ }\href {http://link.aip.org/link/?JAP/58/R1/1} {\bibfield
  {journal} {\bibinfo  {journal} {Journal of Applied Physics}\ }\textbf
  {\bibinfo {volume} {58}},\ \bibinfo {pages} {R1} (\bibinfo {year}
  {1985})}\BibitemShut {NoStop}%
\end{thebibliography}
